\documentclass[%
 reprint,
superscriptaddress,
twocolumn,
 amsmath,amssymb,
 aps,
 prx,
longbibliography
]{revtex4-2}
\usepackage{graphicx}
\usepackage{dcolumn}
\usepackage{bm}
\usepackage{fixmath}
\usepackage[normalem]{ulem}
\usepackage[usenames,dvipsnames]{color}

\usepackage[colorlinks,bookmarks=true,citecolor=blue,linkcolor=blue,urlcolor=blue]{hyperref}

\begin{document}

\preprint{APS/123-QED}

\title{Dynamical Mean Field Theory of Moir\'e Bilayer Transition Metal Dichalcogenides: Phase Diagram, Resistivity, and Quantum Criticality}

\author{Jiawei Zang}
\email{jz3122@columbia.edu}
\affiliation{Department of Physics, Columbia University, 538 W 120th Street, New York, New York 10027, USA}
\author{Jie Wang}
\affiliation{Center for Computational Quantum Physics, Flatiron Institute, 162 5th Avenue, New York, NY 10010, USA}
\author{Jennifer Cano}
\affiliation{Center for Computational Quantum Physics, Flatiron Institute, 162 5th Avenue, New York, NY 10010, USA}
\affiliation{Department of Physics and Astronomy, Stony Brook University, Stony Brook, New York 11974, USA}
\author{Antoine Georges}
\affiliation{Center for Computational Quantum Physics, Flatiron Institute, 162 5th Avenue, New York, NY 10010, USA}
\affiliation{Coll\`ege de France, 11 place Marcelin Berthelot, 75005 Paris, France}
\affiliation{CPHT, CNRS, \'Ecole Polytechnique, Institut Polytechnique de Paris, Route de Saclay, 91128 Palaiseau, France}
\affiliation{DQMP, Universit\'e de Gen\`eve, 24 Quai Ernest Ansermet, CH-1211 Gen\`eve, Switzerland}
\author{Andrew J. Millis}
\email{ajm@columbia.edu}
\affiliation{Department of Physics, Columbia University, 538 W 120th Street, New York, New York 10027, USA}
\affiliation{Center for Computational Quantum Physics, Flatiron Institute, 162 5th Avenue, New York, NY 10010, USA}

\date{\today}

\begin{abstract}
We present a comprehensive dynamical mean field study of the triangular lattice moir\'e Hubbard model, which is believed to represent the physics of moir\'e bilayer transition metal dichalcogenides. In these materials, important aspects of the band structure including the bandwidth and the order and location of van Hove singularities can be tuned by varying the interlayer potential. We present a magnetic and metal-insulator phase diagram and a detailed study of the dependence of the resistivity on temperature, band filling and interlayer potential. We find that transport displays Fermi liquid, strange metal and quantum critical behaviors in distinct regions of the phase diagram. Specifically, we find that the cube-root van Hove singularity  ($\rho(\epsilon) \sim|\epsilon|^{-1 / 3}$) gives a strange metal behavior with a $T$-linear scattering rate and $\omega/T$ scaling. We show how magnetic order affects the resistivity. Our results elucidate the physics of the correlated states and the metal-insulator continuous transition recently observed in twisted homobilayer WSe$_2$ and heterobilayer MoTe$_2$/WSe$_2$ experiments. 
\end{abstract}

\maketitle


\section{\label{introduction} Introduction}

Moir\'e bilayers have emerged as a fascinating system for exploration of correlated electron physics \cite{tbg1,tbg2,ghiotto2021quantum,Wang:2020us,Cornell,cornellWS2WSe2}. Moir\'e bilayers are formed when a monolayer of one material is placed on top of a monolayer of another material (heterobilayer) or when two layers of the same material (homobilayer) are stacked. Bilayers are typically encapsulated with other materials for protection against degradation and to enable electrical contacts. In the heterobilayer case an in-plane moir\'e (very long period) superlattice may be  generated when the two components have a lattice mismatch \cite{Cornell,cornellWS2WSe2,wu2018hubbard}. In the homobilayer case a moir\'e superlattice may be generated when the two layers  are placed at a relative twist angle \cite{ghiotto2021quantum,Wang:2020us,wu2019topological}.  The low energy properties of  moir\'e systems are typically described by few-orbital tight binding-like models in which the single-particle electronic properties  can be tuned over wide ranges by variation of the interlayer potential difference (``displacement field'') while the interaction strength is set by the size of the moir\'e unit cell. The tunability of the relative magnitude of interaction and band scales makes a systematic theory-experiment comparison possible.

Moir\'e bilayers formed from transition metal dichalgocenide (TMDC) materials are now of intense interest \cite{ghiotto2021quantum,Wang:2020us,wu2019topological,Cornell,cornellWS2WSe2,wu2018hubbard,Sarma,ZangHF,Jie_StaggeredField}. In heterobilayer MoTe$_2$/WSe$_2$ the bandwidth can be tuned over about an order of magnitude while in twisted homobilayer WSe$_2$ (tWSe$_2$) properties of the van Hove point including its  Brillouin zone location and the degree of singularity   are controlled by the displacement field \cite{ZangHF,Liang_vhs1,Liang_vhs2}. tWSe$_2$ has recently been found to exhibit a reentrant metal-insulator transition  that is controlled by the carrier concentration and displacement field, with regions of linear-$T$ resistivity near the boundaries to the insulating phases \cite{Wang:2020us,ghiotto2021quantum}. Heterobilayer MoTe$_2$/WSe$_2$ has been reported to exhibit a metal-insulator transition controlled by the displacement field with a scaling collapse of the resistivity data over wide temperature and displacement field ranges \cite{Cornell}. The rich experimental phenomenology combined with the relative simplicity of the systems and the high degree of experimental control suggests that a more comprehensive understanding of the physics of the metal-insulator transitions and the associated strange metal may be possible; this understanding may shed new light on the physics of other strongly correlated systems such as twisted bilayer graphene, high transition temperature cuprates and Sr$_3$Ru$_2$O$_7$ because of van Hove singularity and magnetism \cite{tbglinear,tbglinear2,cuprate,strangemetal}.

The wide range of temperature scales that can be experimentally accessed, from the asymptotic low $T$ regime to temperatures comparable to or greater than bandwidths, provides a new condensed matter approach to the fundamental question of the high temperature physics  of interacting electrons. 
The intermediate and high temperatures regimes ~\cite{highT1,highT2,Palsson_1998,GunnarssonRMP} have been 
mostly experimentally probed in ultracold atom systems ~\cite{Bakr_2019,DeMarco_2019,Thywissen_2019}, while they are generally inaccessible in solid-state systems due to the large  bandwidths. 

In both tWSe$_2$ and MoTe$_2$/WSe$_2$ the low energy physics  is believed to be well modelled by the triangular lattice {\em moir\'e} Hubbard model, which extends  the familiar triangular lattice Hubbard model \cite{wietek2021mott,Tanaskovic,singlesiteDMFT,Goto_2016,PhysRevB.96.205130,PhysRevX.10.021042,chen2021quantum} by tuning important aspects of the band structure and breaking the SU(2) spin symmetry \cite{ghiotto2021quantum, Wang:2020us,Cornell,Sarma,wu2018hubbard,ZangHF}. While the conductivity of interacting lattice models has been analysed \cite{Tanaskovic,Georges_2004,highT1,highT2,McKenzie,Limelette2003,Limelette_ising,Tremblay_ising,Dobrosav_cri,Dobrosav_cri2,DMFTvHs1}, the physics associated with the  moir\'e tunabilty has not been fully explored and the comprehensive analysis needed to compare to current experiments has not been performed.

Here we present a complete phase diagram including  magnetic and metal-insulator transitions and a comprehensive study of the dependence of the resistivity on temperature, band filling and displacement field. We find regions of strange metal behavior as well as Fermi liquid regions and show how magnetic order affects the resistivity and how varying the displacement field leads to critical scaling. In particular, we show that the high order van Hove of degree three, which appears in twisted systems\cite{Liang_vhs1,AndreaTrilayer,guerci2021higherorder,shtyk2017electrons}, provides an alternative route to realize linear resistivity. We present a detailed comparison of our results to experiment.

The rest of this paper is organized as follows. In Section~\ref{sec:Model} we present the model and methods. In Section~\ref{sec:phase} we present the magnetic phase diagram at half filling. In Section~\ref{sec:para} we discuss the resistivity behavior restricted to nonmagnetic solutions
and in Section~\ref{sec:nonfermi} we discuss the linear resistivity associated with the proximity of the van Hove singularity. 
In Section~\ref{sec:mag} we discuss the consequence of the magnetic order in the resistivity. 
In Section~\ref{sec:comparison} we compare our results to experiments, and devote Section ~\ref{sec:summary} to a summary and conclusion. 
Appendices present the details of the band structure and our analytical continuation methods. 

\section{Model and Method \label{sec:Model}}
The  moir\'e Hubbard model is \cite{wu2018hubbard,Sarma,Wang:2020us}:
\begin{align}
	H=-\sum_{\langle i, j\rangle, \sigma=\uparrow,\downarrow}c_{i, \sigma}^{\dagger} t^{i,j}_{\sigma}  c_{j, \sigma}+U \sum_{i} n_{i \uparrow} n_{i \downarrow},
	\label{eq:H}
\end{align}
where $i, j$ represent nearest-neighbor sites on the triangular superlattice, $U$ is an on-site repulsive interaction of magnitude set by the size of the moir\'e unit cell and  $t^{i,j}_{\sigma}$ is the hopping parameter. Longer ranged hopping and interaction terms, neglected in Eq. \ref{eq:H}, are believed not to be of qualitative importance for the carrier concentrations of main interest here, which are near one electron per moir\'e unit cell, although the long ranged interactions may be relevant for Wigner crystal states reported at much lower carrier concentrations \cite{CornellWigner,HarvardWigner}. The moir\'e Hubbard model approximates the physics of the two highest lying valence bands; other bands slightly overlap the moir\'e Hubbard bands at the very lowest band energies, and would have to be considered at very low electron concentrations, but are not relevant to this study which focusses on electron densities near $n=1$ per moir\'e unit cell.

In heterobilayer MoTe$_2$/WSe$_2$, the moir\'e pattern is formed by lattice mismatch. Density functional band theory (DFT) calculations reveal that the hopping parameter is real, spin independent, and tunable over more than an order of magnitude by variations of the displacement field (gate voltage difference between two layers) \cite{wu2018hubbard, Cornell}. In homobilayer tWSe$_2$ the moir\'e pattern is formed by a relative twist between the two component layers. In this paper we focus on the twist angle 4-5$^\circ$ studied in recent experiments. The DFT calculations of Ref.~\cite{Wang:2020us} indicate that for this twist angle the top moir\'e bands are well fitted by the single-orbital tight binding model with hopping magnitude $|t|$ around $\sim 10meV$, and the bands are not topological. Topological bands are found to occur for other twist angles and in other calculations \cite{wu2019topological,Sarma,Futopological}. In the experiments \cite{ghiotto2021quantum, Wang:2020us} the resistivity shows no quantization feature while strange metal and quantum criticality behaviors are observed. Therefore we focus on the topological trivial region in this study. The variation of the displacement field over physically reasonable values changes $|t|$ by about 30\% and more importantly adds a spin-dependent complex phase $t^{i,j}_{\sigma}=|t|e^{i\sigma\phi_{ij}}$ in  the range $0 \lesssim \phi \lesssim \pm \frac{\pi}{3}$.  In Appendix~\ref{appendixA}, we present a brief review of the band structure of tWSe$_2$.

\begin{figure}[b]
	\centering
	\includegraphics[width=1.0\linewidth]{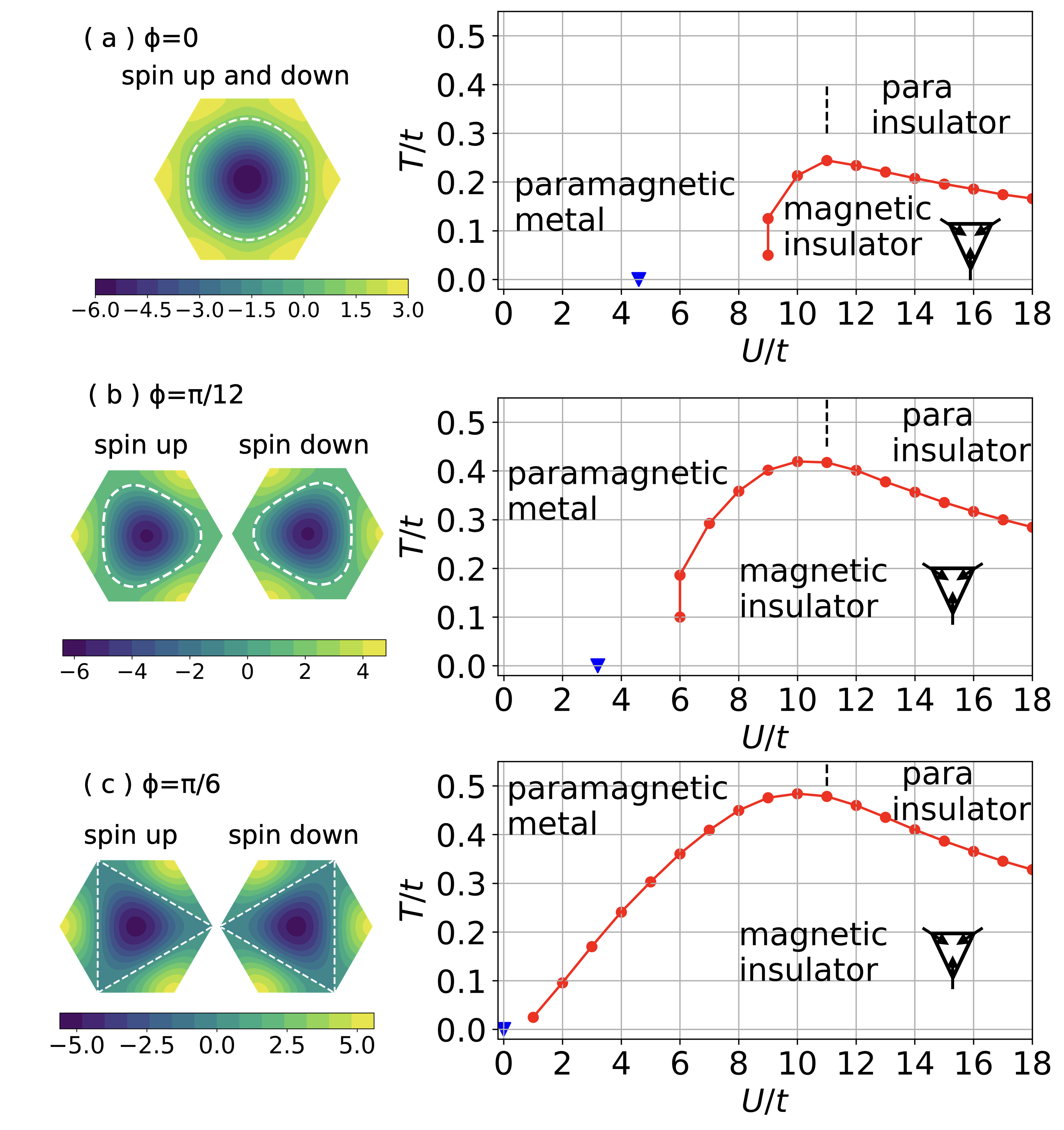}
	\caption{DMFT phase diagram at half filling for (a) $\phi=0$, (b) $\pi/12$ and (c) $\pi/6$. The red lines represent the phase boundary between the magnetic state and the paramagnetic state, determined by the Landau theory. The black dashed line indicates the paramagnetic insulator transition, with resolution $\Delta U=t$, $\Delta T=0.1t$. The blue triangle marks the transition at zero temperature from the Hartree-Fock calculation. On the left we also show the iso-energy dispersion contours at zero interaction, where the dashed lines mark the Fermi surface at half filling.}
	\label{phase}	
\end{figure}

The $\phi$ dependence of the hopping has a dramatic effect on the band structure, as seen in the color maps of constant energy contours shown in the left panels of Fig.~\ref{phase} for $\phi=0, \pi/12$ and $\pi/6$. In these plots the Fermi surfaces corresponding to half filling are indicated as white dashed lines. For a certain value of $\phi$ ($\phi=\pi/6$ in the nearest neighbor model considered here), the  three van-Hove points merge and form a higher order van-Hove singularity in which the density has a power-law divergence $\rho(\epsilon)\sim|\epsilon|^{-1/3}$ \cite{ZangHF,Liang_vhs1,Liang_vhs2}. This higher order van Hove point lies on the Fermi surfaces for half filling with carrier density $n=1$, which are also such that the spin up and spin down Fermi surfaces are nested. 

We use the single-site dynamical mean field theory (DMFT) \cite{RevModPhys.68.13} with the continuous-time hybridization expansion solver (CT-HYB) \cite{RevModPhys.83.349} as implemented in the TRIQS software library \cite{PARCOLLET2015398,Seth_2016}  to perform a comprehensive study of the metal-insulator and magnetic phase diagram, as well as the temperature dependent resistivity, as functions of the carrier concentration, hopping phase $\phi$, relative interaction strength $U/|t|$ and temperature $T/|t|$. The single site DMFT replaces the full momentum and frequency dependent self energy by a self energy that depends only on frequency. Within this approximation quantum and thermal fluctuations are treated  on an equal footing, and  a non-perturbative treatment of the physics is possible. 

\section{Phase Diagram at half filling}\label{sec:phase}
In Fig.~\ref{phase}, we present the computed phase diagram at carrier concentration $n=1$. Our results for $\phi=0$ are consistent with previous DMFT work \cite{wietek2021mott,Tanaskovic,Goto_2016,singlesiteDMFT,PhysRevLett.93.016406}. The low $T$, small $U$ phase is a Fermi liquid metal;  as $U/t$ is increased beyond $U_{c}\approx 9t$ at low $T$,  a first-order transition to a $120^\circ$ magnetically ordered insulating state occurs (spin pattern shown on figure). For large and small $U$ the phases found in DMFT agree with the phases found in other work, but it is likely that for a relatively narrow range of  $U\sim 9t$ a different insulating phase occurs, possibly a spin liquid \cite{ PhysRevB.96.205130,PhysRevX.10.021042,chen2021quantum}. This phase is not captured by the single-site DMFT formalism and won't be discussed in this paper. 

Within the DMFT approximation at $\phi=0$ and at low $T$ the magnetic transition temperature jumps discontinuously (within our resolution) to about $0.125t$ at $U_c\approx9t$ and then varies non-monotonically on further increase of $U$. For $U > 9t$ the temperature driven magnetic transition is second order. As $U$ is varied at higher $T$ a paramagnetic metal to paramagnetic insulator crossover is found at $U\sim 11t$ indicated by the vertical dashed line in Fig.~\ref{phase}.  Within our resolution ($\Delta U=t$, $\Delta T=0.1t$)  the crossover is essentially independent of  $\phi$.  If magnetism is suppressed, then as $U$ is increased at very low $T$, single-site DMFT finds a first order  nonmagnetic metal-nonmagnetic insulator transition \cite{KotliarDMFT} with a critical end point at $T_c^p\approx0.1t$, $U_{c}^p\approx11t$ with interesting scaling associated with the endpoint of the paramagnetic metal-paramagnetic insulator transition \cite{Limelette_ising,PRLLimelette}. For the triangular lattice model considered here the critical end-point is well within the magnetic phase, and will not be discussed further here.

As $\phi$ is increased, the high $T$ paramagnetic metal-paramagnetic insulator crossover does not change. However, the magnetic phase expands in both temperature and $U$ and at $\phi=\pi/6$ the magnetic phase extends to $U=0$ and the transition is everywhere second order.  The $\phi$-dependence of the magnetic phase boundaries can be understood from the fermiology, shown in the left panel of Fig.~\ref{phase}: as $\phi\rightarrow \pi/6$ the Fermi surface becomes nested and the van Hove singularities merge to form a third order singularities at $K/K^\prime$ points of the Brillouin zone. 

The ability of DMFT to capture quantum  fluctuations leads to differences from our previous $T=0$ Hartree-Fock phase boundary \cite{ZangHF}, shown as blue triangles in the figure. The transition temperature is also greatly overestimated by the temperature dependent Hartree-Fock approximation. Of course, DMFT cannot capture the long wavelength physics associated with the true Kosterlitz-Thouless transition critical behavior implied by the $xy$ symmetry of the magnetic action \cite{ZangHF}, but the computed transition temperature is believed to adequately reflect the crossover scale below which magnetic correlations become very long ranged \cite{wietek2021mott}.

\section{Resistivity in the paramagnetic state}\label{sec:para}
The resistivity is perhaps the most experimentally accessible observable in twisted bilayer systems. In this section and Section~\ref{sec:nonfermi} we focus on the resistivity of the paramagnetic phase, turning to the effects of magnetism in Section~\ref{sec:mag}. The conductivity of the Hubbard model has been studied previously \cite{Tanaskovic,Georges_2004,highT1,highT2,McKenzie,Limelette_ising,Tremblay_ising,Dobrosav_cri,Dobrosav_cri2,DMFTvHs1}. The published conductivities exhibit Fermi liquid metal and ``Mott insulating'' regimes, depending on carrier concentration and interaction strength. An interesting scaling behavior has been found to be associated with the metal-insulator transition. The critical endpoint of the paramagnetic metal-paramagnetic insulator first order metal-insulator transition line gives rise to an Ising criticality ~\cite{Rozenberg_1999,Limelette_ising,Georges_2004,Tremblay_ising} while over a much broader range at high $T$ a data collapse associated with a critical scaling has been reported \cite{Dobrosav_cri,Dobrosav_cri2}. Here we investigate how the displacement field affects the conductivity, with a particular focus on the tunable van Hove singularity.

In the dynamical mean field approximation and in the paramagnetic phase the conductivity is given by \cite{RevModPhys.68.13}:
\begin{equation}
    \sigma_{\rm DC}^{\rm non-mag}=\frac{2\pi e^2}{\hbar S N } \sum_{\mathbf{k}}(\frac{\partial \epsilon_\mathbf{k}}{\partial k_{x}})^{2}\int d\omega A(\epsilon_k, \omega)^{2}\frac{-\partial n_F(\omega)}{\partial\omega},\\
\end{equation}
where the $\mathbf{k}$ sum is over the Brillouin zone of the triangular lattice with $N$ sites, $S=\sqrt{3}a^2/2$ is the area of the unit cell, $\epsilon_k$ is the dispersion, $n_F(\omega)$ is the Fermi-Dirac function, and $A(\epsilon_k, \omega)=-\frac{1}{\pi} \operatorname{Im}[\omega+\mu-\epsilon_k-\Sigma(\omega)]^{-1}$ is the spectral function. The resistivity requires knowledge of the self energy $\Sigma(\omega)$ for real frequencies; we obtain $\Sigma(\omega)$ via   maximum entropy analytical continuation (\emph{MaxEnt}) \cite{JARRELL1996133,MaxE_triqs} of our Quantum Monte Carlo (QMC) data, and spot check the results both by analysing the data directly on the imaginary axis and by comparison to the Pad\'e method of continuation \cite{Pade}.

\begin{figure}[t]
	\centering
	\includegraphics[width=1.0\linewidth]{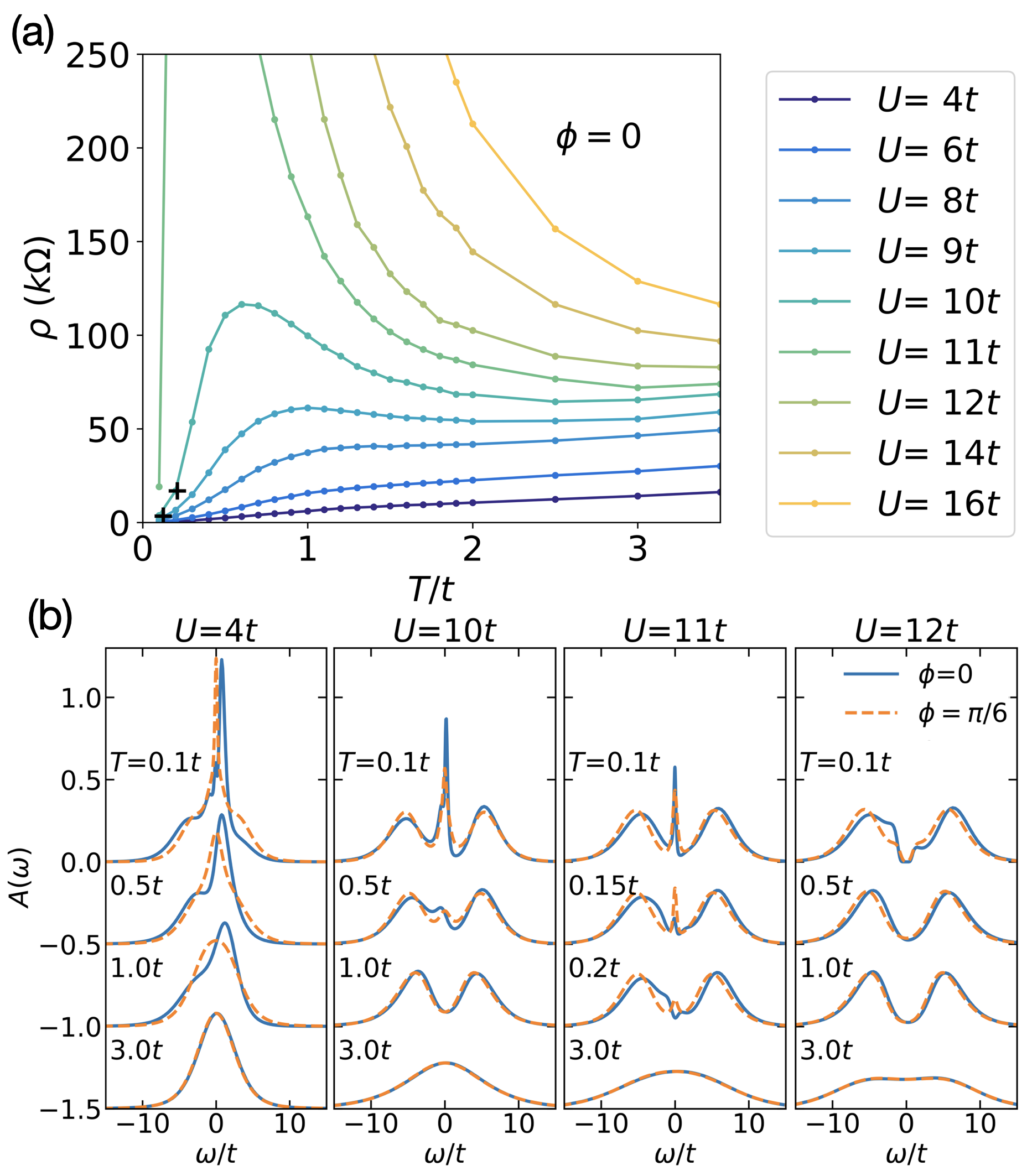}
	\caption{(a): Temperature dependence of resistivity at half filling for $\phi$=0 with different interactions. The black cross marks the magnetic transition temperature for $U/t=9,10$. (b): Spectral function for $\phi=0$ (solid line) and $\pi/6$ (dashed line) at representative temperatures for different interactions.}
	\label{res0}	
\end{figure}

Fig.~\ref{res0}~(a) shows the resistivity calculated at half filling and $\phi=0$ over a wide range of temperatures and interactions. These results agree with previous calculations \cite{Tanaskovic,Georges_2004,highT1,highT2,McKenzie,Limelette_ising,Tremblay_ising,Dobrosav_cri,Dobrosav_cri2,DMFTvHs1}. Three temperature regimes are evident: a low T regime ( $T\lesssim 0.4t$) where an asymptotic low $T$ behavior is evident (either gapped, for large $U$, or tending to perfect conduction for smaller $U$), a high T regime (roughly $T\gtrsim 2t$ with the precise boundary being $U$-dependent and becoming larger than the largest scale shown for $U\gtrsim 12t$) where the resistivity is approximately of the form $\rho=A+BT$ \cite{highT2}, and an intermediate temperature crossover regime. As shown in Fig.~\ref{res0}~(b) (solid line), the different regimes are associated with different behaviors of the electronic spectral function, with the low $T$ behavior associated with the presence either of a gap or a sharp quasiparticle peak, the high $T$ behavior associated with a very broad spectral function with weak temperature dependence arising only from the $T$ dependence of the chemical potential, as previously noted in Refs.~(\onlinecite{Palsson_1998,highT2}).

\begin{figure}[t]
	\centering
	\includegraphics[width=0.9\linewidth]{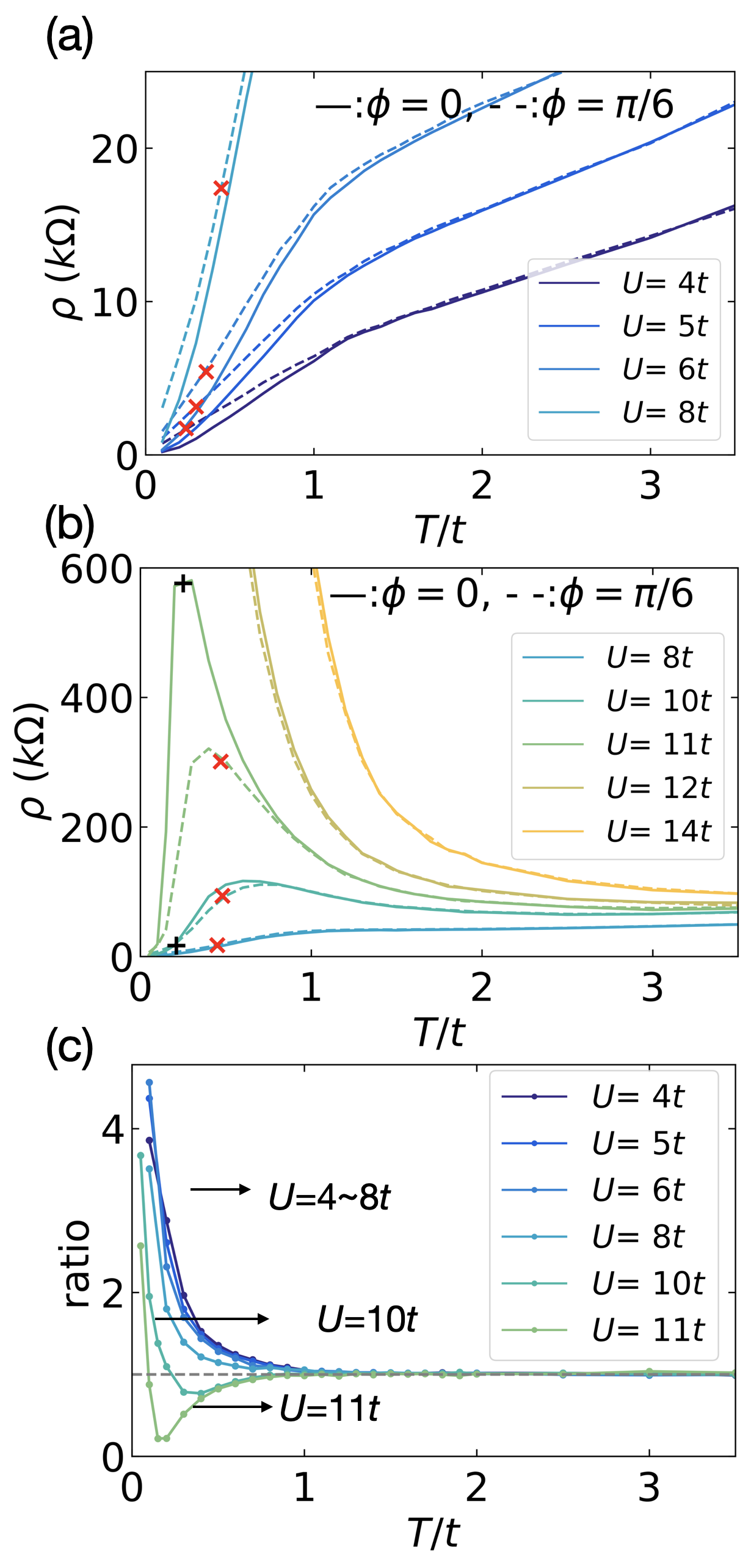}
	\caption{(a) and (b): Resistivity for $\phi=0$ (solid line) and  $\phi=\pi/6$ (dashed line) for moderate and large interactions, respectively. The DMFT computation is restricted to nonmagnetic solutions, and the black ($\phi=0$) and red ($\phi=\pi/6$) crosses mark the magnetic transition temperature. (c): The ratio of resistivity of two phases $\rho(\phi=\pi/6)/\rho(\phi=0)$ at different temperature and interactions.}
	\label{para_res}	
\end{figure}

We now proceed to the new features at $\phi\neq0$. Fig.~\ref{para_res} compares the resistivity at $\phi=0$ to the resistivity at $\phi=\pi/6$. Panel (a) highlights the moderate-interaction regime where the paramagnetic phase  ground state is a metal. We see that varying the phase angle does not change the resistivity significantly at higher $T \gtrsim t$, but has a profound effect at lower $T$, changing the scaling with temperature from quadratic to linear and concomitantly increasing the magnitude. Panel (b) presents the large $U$ metal-insulator crossover regime. We see that changing $\phi$ again has essentially no effect at higher temperatures, with in particular the Mott gap insensitive to the displacement field. However, in the crossover regime $U\approx 10-11t$, increasing $\phi$ has a much stronger effect. As $T$ is decreased below $T\sim t$ into the metal insulator crossover regime, the resistivity is governed by two competing effects characterized by different energy scales. A Hubbard gap structure (visible in the $U=10,11t$ spectral function shown in Fig.~\ref{res0}(b)) begins to open, causing a rapid upturn in the resistivity; then at lower $T$ a mid gap quasiparticle peak appears \cite{RevModPhys.68.13}, leading to a rapid downturn in the resistivity and ultimately metallic behavior. We see that the initial upturn is essentially the same for $\phi=0$ and $\phi=\pi/6$, consistent with a minimal effect of $\phi$ on the Mott gap. However, the difference in temperature of the resistivity maximum shows that the enhanced density of states related to the Fermi surface van Hove singularity at $\phi=\pi/6$ favors the formation of the mid-gap quasiparticle peak (seen also in Fig.~\ref{res0}(b) at $U=11t$), leading to the much lower resistivity at $T \sim 0.5t$. Finally, at very low $T$, the enhanced scattering arising from the van Hove singularity makes the resistivity at $\phi=\pi/6$ again larger. The non-monotonic behavior is highlighted in panel (c) which presents the ratio $\rho(\phi=\pi/6)/\rho(\phi=0)$. The $\phi$ dependence of the metal-insulator crossover may be expected to lead to a $\phi$-dependence of the location of the critical end-point of the metal-insulator transition  and may affect other features of the metal-insulator crossover line and the critical end point. The $\phi$ dependence is weak, essentially outside of our resolution $\Delta U\sim t$, $\Delta t\sim 0.1t$. A detailed study of these aspects is left for the future.

\section{Linear resistivity at \texorpdfstring{$\phi=\pi/6$}{phi=pi/6}}\label{sec:nonfermi}

Panel (a) of Fig.~\ref{para_res} shows that at $\phi=\pi/6$, where the high order van Hove singularity is at the Fermi surface, the asymptotic low temperature resistivity is $\rho \sim T$.  At very low temperature, the resistivity is related to the scattering rate, or the imaginary part of the self energy Im$\Sigma(\omega)$, and it is important to verify that the linearity we find is not an artifact of the maximum entropy analytic continuation method used to obtain most of our resistivity data. 

As a first check we examine directly the imaginary axis behavior. In the Fermi liquid regime at sufficiently low $T$, the self-energy on the Matsubara-axis has the low-energy expansion:
\begin{equation}
\operatorname{Im} \Sigma(i \omega_{n}, T) \approx(1-\frac{1}{Z}) \omega_{n}+\frac{\omega_{n}^{2}-(\pi T)^{2}}{E} + \mathcal{O}(\omega_{n}^{3}),\label{eq:sigma}
\end{equation}
which implies at the first Matsubara point $i\omega_{n=0}=i\pi T$,
\begin{equation}
\frac{\operatorname{Im} \Sigma(i  \pi T,T)}{T}=(1-\frac{1}{Z}) \pi + \mathcal{O}(T). \label{eq:sigma2}
\end{equation}

So, a deviation from linearity with $T$ of the self energy at the lowest Matsubara frequency indicates a non-Fermi-liquid behavior \cite{EAKimlinearR,Chubukov_2012}. 

To understand what the deviations might be we observe that the real and imaginary self energies  are connected by the Kramers–Kronig relation:
\begin{equation}
\operatorname{Im} \Sigma(i\omega_{n}, T)
 =-\frac{\omega_{n}}{\pi}\int^{\infty}_{-\infty}d\omega\frac{\text{Im}\Sigma (\omega,T)}{\omega^2+\omega_n^2}.\label{eq:KK}
\end{equation}

\begin{figure}[htbp]
	\centering
	\includegraphics[width=1.0\linewidth]{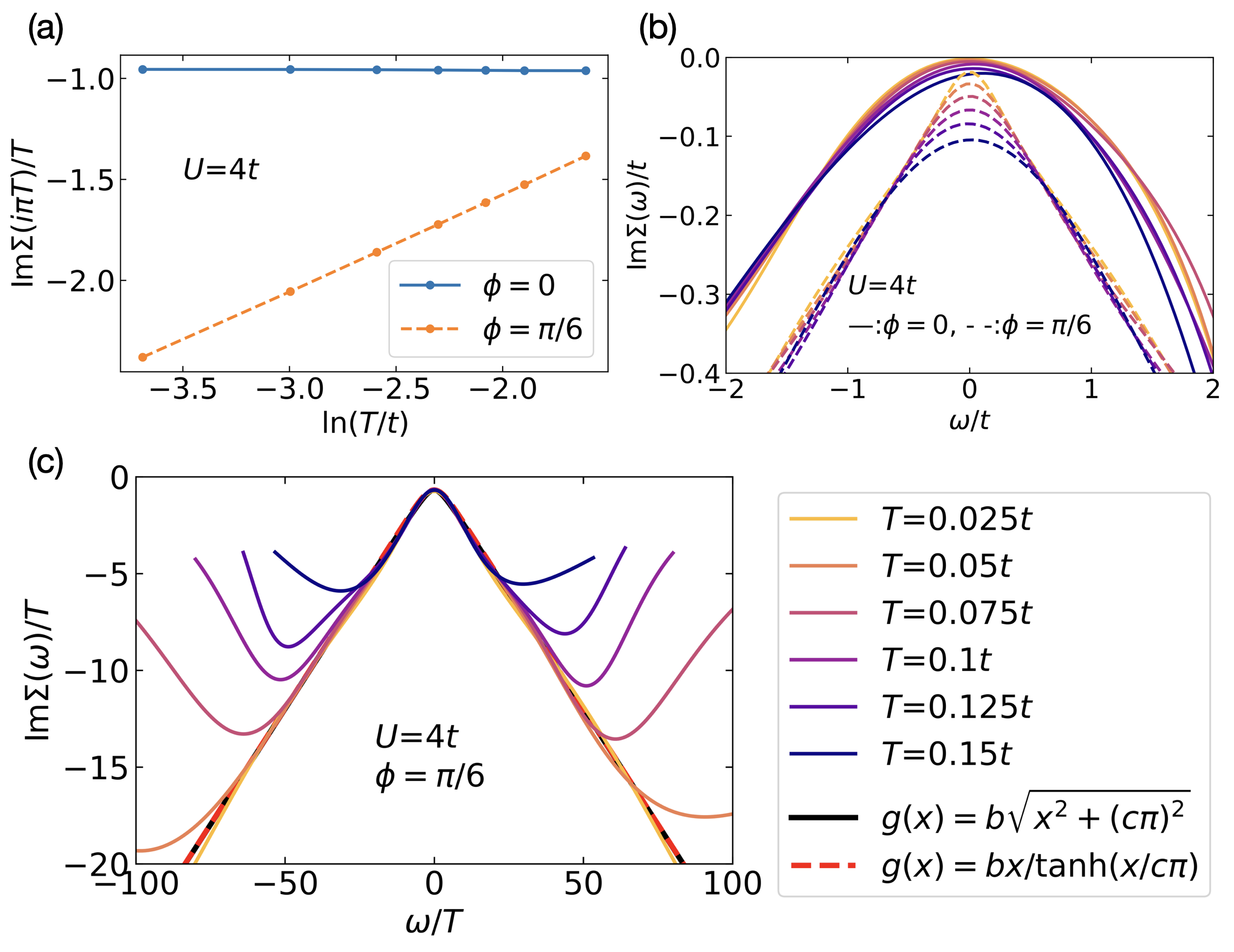}
	\caption{(a): Im$\Sigma(i \omega_{n}=i\pi T, T)/T$ as a function of ln$(T/t)$ at the first Matsubara frequency point. (b): The imaginary part of the self energy Im[$\Sigma(\omega)$] for $\phi=0$ and $\pi/6$ at $U=4t$ and half filling. (c): Im[$\Sigma(\omega)$]/$T$ versus $\omega/T$ for $\phi=\pi/6$.  The black and red lines are reference lines $g(x)=b\sqrt{x^2+(c\pi)^2}$ and $g(x)=bx/\tanh(x/(c\pi))$, correspondingly. The slope $b\approx0.24$ is obtained independently from the fitting shown in Panel (a) and $c\approx 0.92$ is obtained from the linear fitting of Im[$\Sigma(\omega=0)$].}
	\label{sigma}
\end{figure}

A natural guess for the non-Fermi liquid behavior is a scaling function such as the ``marginal Fermi liquid" form, where Im$\Sigma(\omega)= b(\omega^2+(c\pi T)^2)^{1/2}$ \cite{CMVarma}. Inserting this expression into the Kramers–Kronig relation gives
\begin{eqnarray}
\frac{\operatorname{Im} \Sigma (i\pi T, T)}{T}
 =-2b~\ln\frac{E^\star}{T} + \mathcal{O}(T),
\label{eq:sigma4}
\end{eqnarray} with $E^\star$ a cutoff energy scale.
With this idea, we plot Im$\Sigma(i\pi T, T)/T$ as a function of $\ln T$ in Panel (b) of Fig.~\ref{sigma}. For $\phi=0$, Im$\Sigma(i\pi T, T)/T$ is a constant, confirming the Fermi liquid behavior; while for $\phi=\pi/6$, Im$\Sigma(i\pi T, T)/T$ is proportional to $\ln T$ with slope $b=0.240\pm0.001$, confirming the non-Fermi liquid behavior.  

Next, we employ a longer quantum Monte Carlo loop $N_{\rm QMC}=10^{10}$ to obtain  higher accuracy data for $\Sigma(i\omega_n)$ at $U=4t$, $\phi=0$ and $\pi/6$ and several temperatures. We  analytically continue these data using the Pad\'e method \cite{Pade} which  is believed to provide a more accurate estimate of the small frequency and low temperature self energy than the maximum entropy method. Panel (b) of Fig.~\ref{sigma} shows that for $\phi=0$ we obtain the quadratic behavior expected in a Fermi liquid while for $\phi=\pi/6$ we see a linear $\omega$ dependence  at $T\ll\omega$ and a linear $T$ dependence at $\omega\ll T$. To further analyse this behavior,  we replot the data as Im$\Sigma(\omega)/T$ against $\omega/T$, as shown in Fig.~\ref{sigma}(c). We see that this scaling of the variables leads to an essentially perfect scaling collapse, implying that the low energy self energy is a scaling function Im$\Sigma(\omega,T)\rightarrow T g\left(\frac{\omega}{T}\right)$ with $\text{lim}_{x\rightarrow\infty} g(x)=bx$ and $\text{lim}_{x\rightarrow 0} g(x)=bc\pi$. 
Deviations  occur when $\omega$ is comparable to $U$. %
We compare our results to two specific forms of the scaling function: 
$g(x)=b\sqrt{x^2+(c\pi)^2}$ and $g(x)=bx/\tanh(x/c\pi)$. 
The first one corresponds to the ``marginal Fermi liquid" form  \cite{CMVarma} and the second one is inspired by 
the one found in conformally invariant quantum impurity models~\cite{Antoine_sigma} (for which, however, $c=2/\pi$). The slope $b$ used in the figure is directly obtained from the fitting of Matsubara frequency and it agrees well with the Pad\'e data. $c\approx0.92$ is obtained from the linear fitting of Im[$\Sigma(\omega)$] at $\omega=0$ but this value is sensitive to the uncertainties of analytical continuation. Determining the precise scaling function associated with the self energy is left for future work.

In summary, we analytically and numerically confirmed that at $\phi=\pi/6$ the low temperature resistivity in the metallic phase is linear. The linearity is theoretically expected for a van Hove singularity of degree three within the DMFT approximation \cite{DMFTvHs1,DMFTvHs2}. The single-site DMFT is a local approximation; study of the self energy \cite{Sr2RuO4,RiceLaSrCuO,strangemetal} and resistivity \cite{Tanaskovic,KotliarDMFT} beyond the local self energy approximation   are left to future work. 

\begin{figure}[htbp]
	\centering
	\includegraphics[width=1.0\linewidth]{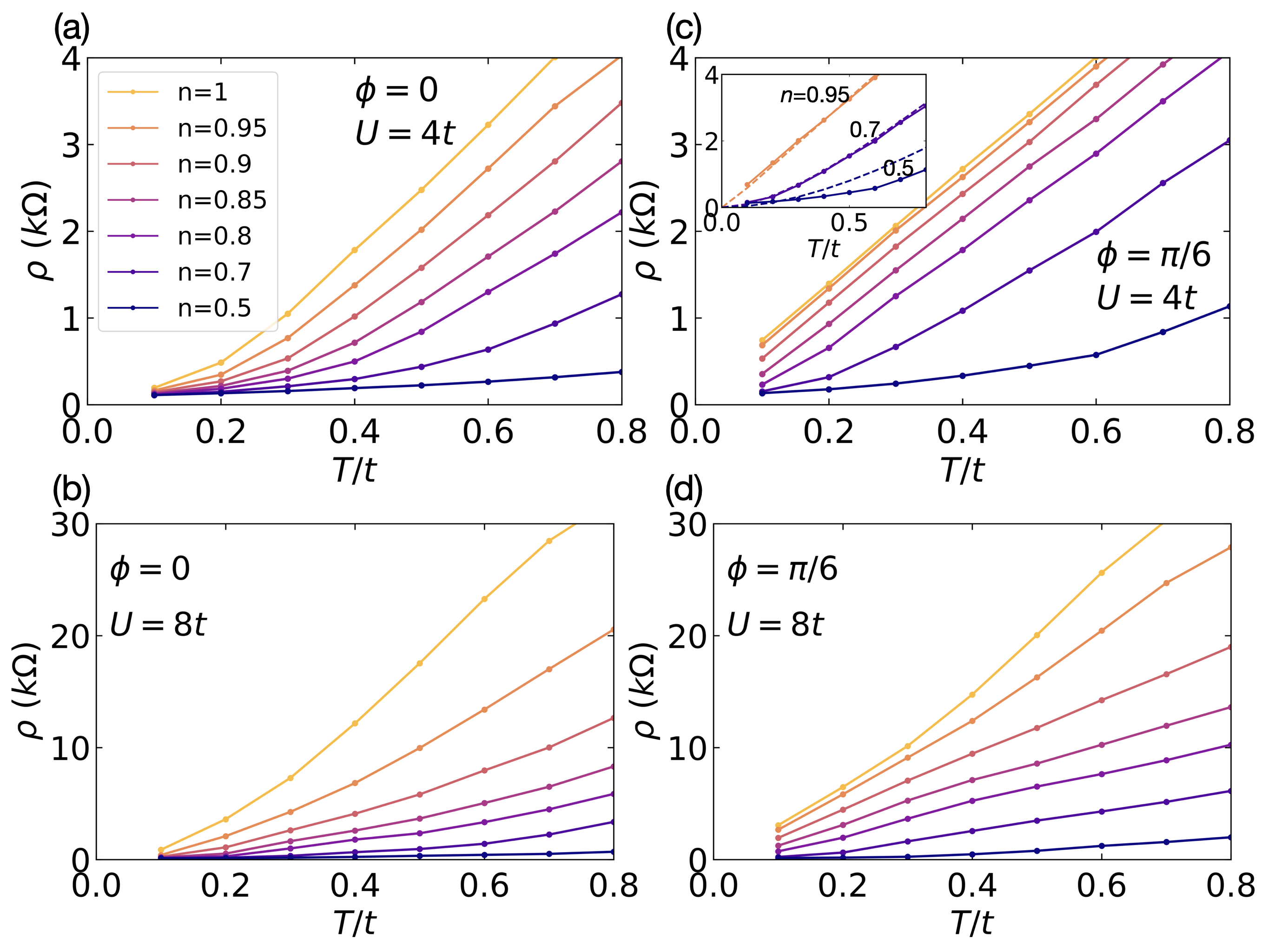}
	\caption{Resistivity from paramagnetic DMFT solution at $U/t=4,8$ for $\phi=0,\pi/6$ for different carrier concentrations, shown in legend. The inset of Panel (c) shows resistivity data (solid line) and dashed curves calculated by a simple formula $\rho=\frac{6.5T^2}{T+T^\star}$ with the crossover scale $T^\star=6t(1-n)^2$.}
	\label{para_dope}
\end{figure}

Finally,  we consider the doping-driven crossover of the metallic resistivity from the $T$-linear behavior found when the Fermi surface touches the third order van Hove point to the $T^2$ behavior expected when the van Hove point is not at the Fermi surface. Fig.~\ref{para_dope} shows the $T$-dependence of the low-$T$ resistivity computed for  two representative interactions $U=4t$ and $8t$ for different hole dopings. For $\phi=0$, as expected, the system has a clear $T^2$ dependence at low temperature at all dopings with a dramatic increase in coefficient as $n\rightarrow 1$. For $\phi=\pi/6$, the low T temperature dependence gradually changes from $T-$linear to $T^2 $ as doping is increased, but the magnitude of the resistivity is less sensitive to the carrier concentration near $n=1$. As shown in the inset of Fig.~\ref{para_dope} (c), at $U=4t$ and $\phi=\pi/6$, we find that the crossover from the $T$-linear to $T^2$ resistivity is approximately given by $\rho\sim\frac{6.5T^2}{T+T^\star}$ with the crossover scale $T^\star=6t(1-n)^2$. As seen the formula describes the data well for $n\gtrsim0.7$ and T $\lesssim0.6t$. The crossover scale $T^\star$ becomes very small for $n$ near $1$.

\section{Resistivity in the magnetic state}\label{sec:mag}

\begin{figure}[htbp]
	\centering
	\includegraphics[width=1.0\linewidth]{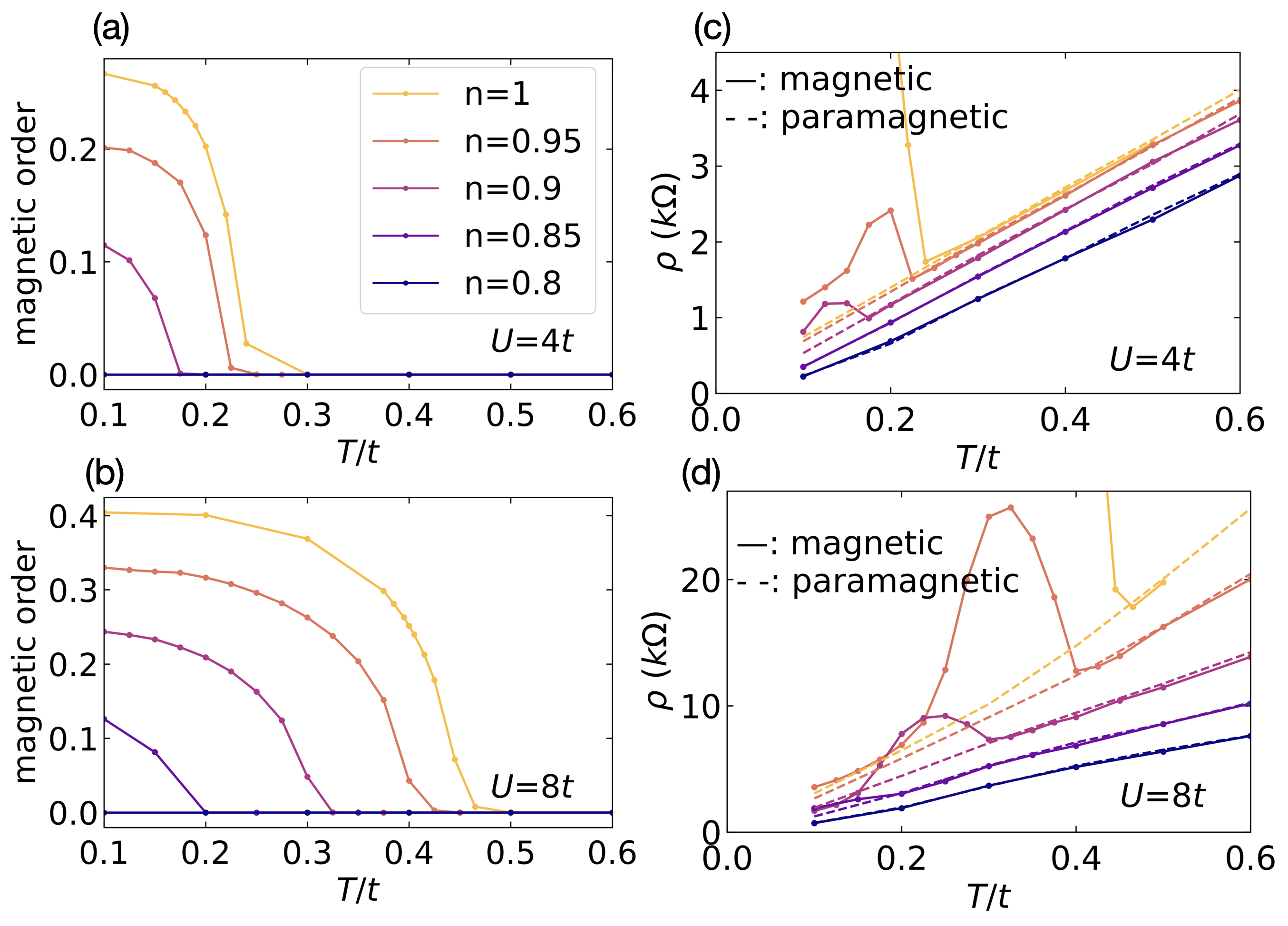}
	\caption{(a) and (b): Amplitude of the magnetic order (Eq.~\ref{eq:mdef} for $U/t=4,8$ and densities shown, at $\phi=\pi/6$. (c) and (d): Solid line: resistivity computed at densities shown  for $U/t=4,8$, including the effects of magnetic order. Dashed line: resistivity, computed in the paramagnetic state. }
	\label{mag_res}
\end{figure}

We next explore the consequence of the magnetic order for the resistivity, focusing on $\phi=\pi/6$. We consider two representative interaction strengths, $U=4t$ and $U=8t$ and consider the temperature and doping dependence of the magnetic order parameter and the resistivity. In computing the conductivity we note that the magnetic order breaks the translation symmetry so the calculation is formulated in the reduced magnetic Brillouin zone. However, we showed previously that a space-dependent spin rotation combined with a spin dependent gauge transformation maps the 120$^\circ$ antiferromagnetic state into an in-plane ferromagnetic state with hopping phase $\phi=\pi/2$ \cite{ZangHF}. In this transformed frame the magnetic order parameter is:
\begin{equation}
    m=\text{Tr}\left(\sigma_x \mathbf{G}\right),\label{eq:mdef}
\end{equation}
where $\mathbf{G}_{\sigma\sigma'}$ is the Green function matrix in the spin basis. The DC conductivity is:
\begin{equation}
    \sigma_{\rm DC}^{\rm mag}=\frac{\pi e^2}{\hbar SN}\sum_{\mathbf{k},\sigma,\sigma^\prime}v^x_{\mathbf{k}\sigma}v^x_{\mathbf{k}\sigma^\prime}\int d\omega A_{\mathbf{k}\sigma\sigma^\prime}(\omega)A_{\mathbf{k}\sigma^\prime\sigma}(\omega)\frac{-\partial n_F(\nu)}{\partial\omega},
\end{equation}
where $v^x_{\mathbf{k}\sigma}=\frac{\partial \epsilon_{\mathbf{k}\sigma}}{\partial k_{x}}$ is the group velocity and $A_{\mathbf k\sigma\sigma'}(\omega)=-\frac{1}{\pi} \operatorname{Im} G_{\sigma\sigma'}(\mathbf{k}, \omega)$ is the spectral function, which now has spin off-diagonal components because of the in-plane magnetic order.

The left panels of Fig.~\ref{mag_res} show the magnetic order parameter for different carrier concentrations. We see that the ordering temperature and amplitude of magnetic order decrease as the system is doped away from the half filled point. The right panels show the corresponding temperature dependent resistivities. We see that the onset of magnetism leads to an abrupt increase in the resistivity. For $n=1$, the system is fully gapped and the resistivity in the magnetic state diverges at $T=0$. For $n\neq 1$, the Fermi surface is not fully gapped and after the initial increase the resistivity again decreases as temperature is decreased. At temperature below but not too far below the ordering temperature, the resistivity is higher in the magnetic state than it would be if the magnetization were suppressed. It might be that at very low $T$ the fact that the magnetization gaps out some of the fluctuations that scatter electrons means that the resistivity of the magnetic state may even become lower than that of the paramagnetic state as $T\rightarrow0$.
\begin{figure}[htbp]
	\centering
	\includegraphics[width=1.0\linewidth]{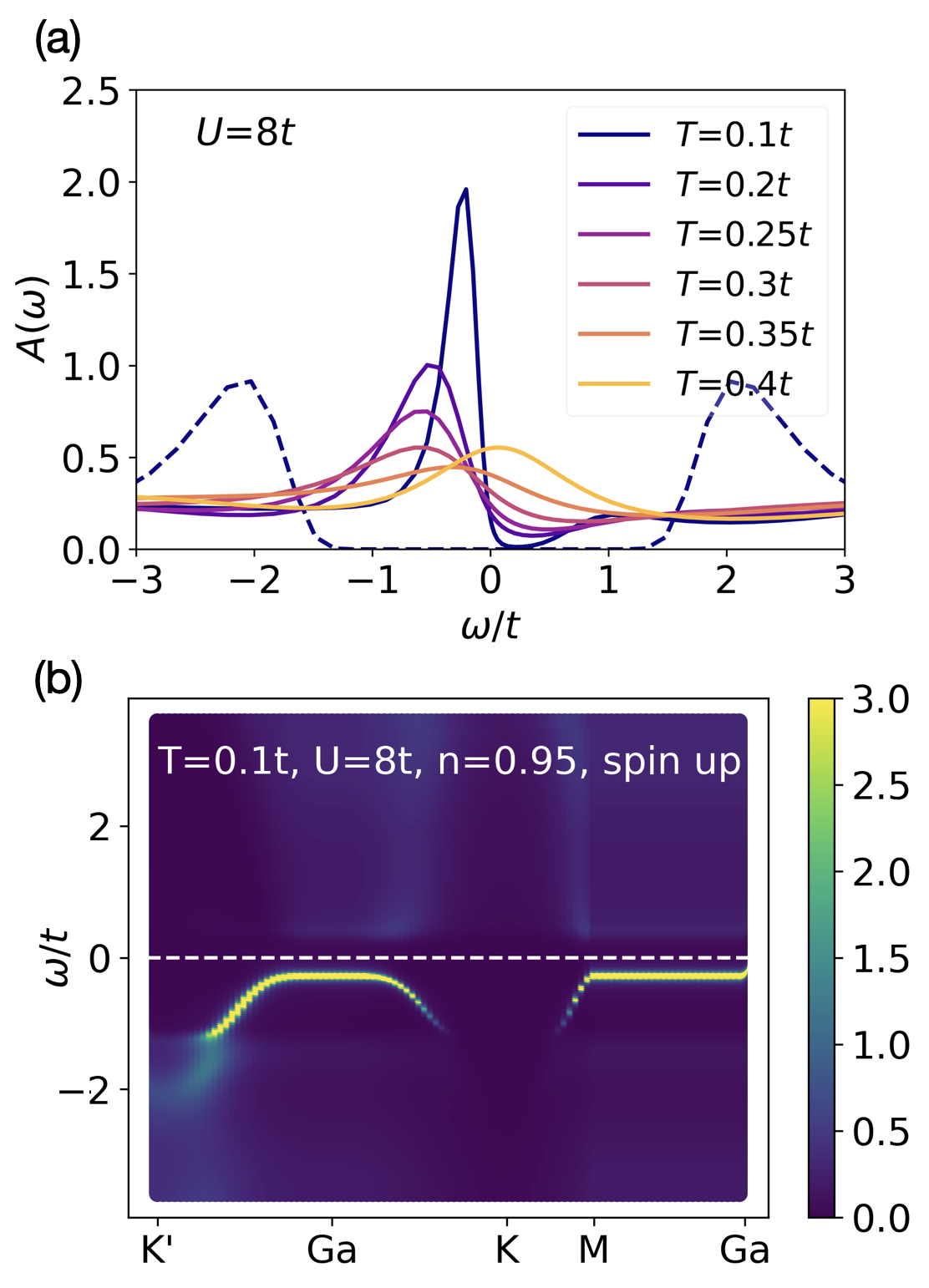}
	\caption{(a): The momentum integrated spectral function for spin up $A_{\uparrow\uparrow}(\omega)$ at $U=8t$ and carrier concentration $n=0.95$ is plotted at $\phi=\pi/6$ and different temperatures (solid line). For comparison, the spectral function for $n=1$ at $T=0.1t$ is shown as  the dashed line. (b): The momentum dependent spectral function $A_{\bm k\uparrow\uparrow}(\omega)$ for spin up along a momentum line at $U/t=8$, $n=0.95$ and $\phi=\pi/6$. The spectral function is analytically continued from the self energy $\Sigma(\omega)$ using \emph{MaxEnt}. The $\omega=0$ (dashed line) corresponds to the Fermi level.}
	\label{spectral}
\end{figure}

The detailed temperature and doping dependence of the conductivity is controlled in a subtle way by the evolution of the chemical potential and the magnetic order parameter. Panel (a) of Fig.~\ref{spectral} shows the momentum-integrated spectral function for $n=0.95$ and $U=8t$. At zero temperature (not shown), the Fermi level (here defined to be at $\omega=0$) is just below the upper edge of the valence band, leading to a hole pocket (of very elongated aspect ratio because of the perfect nesting at $n=1$). As temperature increases, two effects compete with each other. On the one hand, the chemical potential moves rapidly into the gap due to the thermal effects. This is seen in Fig.~\ref{spectral}~(a) as a downward shift in the peak of the spectral function in the range $0.1t\lesssim T\lesssim 0.3t$  (recall that $\omega=0$ defines the chemical potential) and in the momentum resolved spectral function shown in panel (b). This effect increases the resistivity.  On the other hand, the magnetic order gradually disappears leading to a gap decrease, that decreases the resistivity. As temperature further increases, the gap closes and the peak in the spectral function moves back to $\omega\approx 0$. Therefore the resistivity goes back to the paramagnetic behavior.

\section{Comparison to experiments}\label{sec:comparison}

Here we compare our theoretical results to recent experiments \cite{ghiotto2021quantum,Wang:2020us,Cornell} on moir\'e bilayer TMDC materials. Ghiotto and co-workers \cite{ghiotto2021quantum} study tWSe$_2$, finding a reentrant  metal-insulator transition tuned by carrier concentration and displacement field with an interesting scaling of the temperature dependence of the resistivity near the metal-insulator transition point. Li and co-workers \cite{Cornell} study heterobilayer MoTe$_2$/WSe$_2$, where varying the displacement field drives a metal to insulator transition with scaling exhibited over a wide temperature range. All of these experimental phenomena are also found in our calculations; we present here a more detailed discussion of the correspondence between theory and experiment.

\subsection{tWSe\texorpdfstring{$_2$}{2}}

As noted above,  density functional band calculations \cite{Wang:2020us} performed for the system studied by Ghiotto and co-workers are consistent with a basic hopping parameter $t\sim 10$meV which increases moderately as the displacement field is increased and a phase angle that can be tuned from $\phi=0$ to $\phi\approx \pi/3$ over the physically relevant displacement field range. The displacement field dependence of $t$ and $\phi$ is given in  Appendix \ref{appendixA}.

\begin{figure}[htbp]
	\centering
	\includegraphics[width=1.0\linewidth]{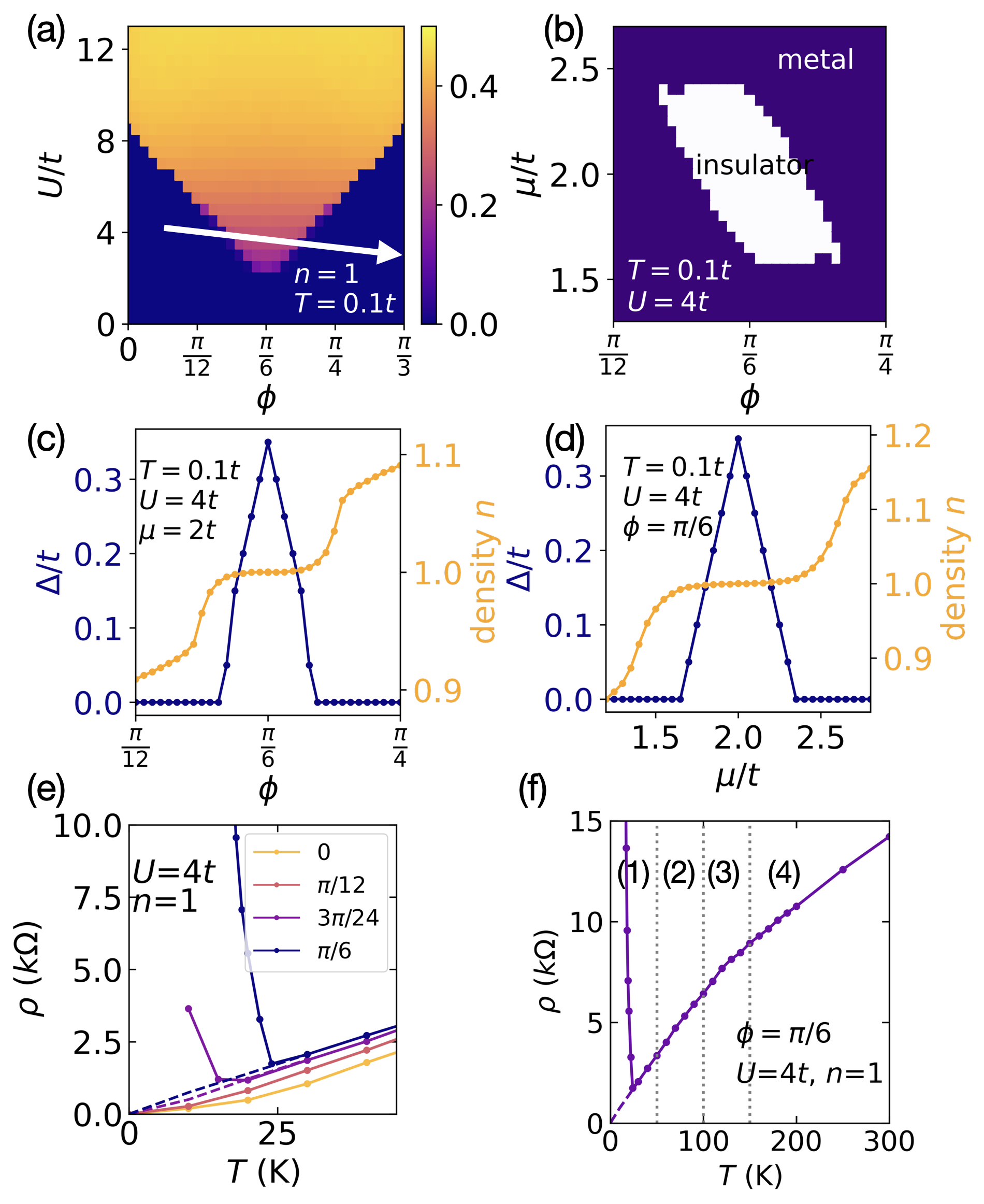}
	\caption{(a) Magnetization as a function of phase $\phi$ and interaction $U$ at half filling. The color represents the magnetization amplitude. The white arrow shows the parameter space trajectory followed when the displacement field is increased in experiments. (b) Metal-insulater reentrant behavior as a function of chemical potential $\mu$ and phase $\phi$ at $U=4t$. We numerically determine the insulating state: the system is insulating if density remains at half filling ($\Delta n<0.01$) with varied chemical potential and if magnetization $m>0.1$. (c) Gap size $\Delta$ and density $n$ verse $\phi$ at $\mu=2t$. (d) Gap size $\Delta$ and density $n$ verse $\mu$ at $\phi=\pi/6$. The gap size is extracted from panel (b), which is the energy difference between $\mu$ and the closest band. (e,f) Resistivity for different phase $\phi$ at half filling. The dashed line shows the resistivity computed in the paramagnetic state. In the figure we choose $t$ =100$K$. In (f) we mark 4 regions that will be explained in the main text.}
	\label{phaseWSe2}
\end{figure}

The key experimental observation is a low $T$ insulating state that exists only in a narrow range of  displacement field, so that in particular by varying the displacement field at fixed carrier concentration the system can be tuned from metal to insulator back to metal. To make a direct comparison, we plot the phase diagram as a function of interaction and phase $\phi$ at half filling and $T=0.1t$, as shown in Fig.~\ref{phaseWSe2} (a). For U $\lesssim 8t$ the properties are reentrant as $\phi$ is varied, with a metallic phase at $\phi$ = 0 giving way to an insulating phase for $\phi$ near $\pi$/6 then evolving back to a metallic phase as $\phi$ is further increased, which is consistent with the experimental findings. Our theoretically calculated insulating state is due to magnetic order, which occurs for displacement fields such that the higher order van Hove singularity is near the Fermi surface. The association of insulating behavior with magnetic order is strengthened by the experimental observation of a sign change of the Hall coefficient as the insulating state is approached by varying doping. This sign change is consistent with the Fermi surface reconstruction expected from magnetic order. The magnetic order we find is an in-plane magnetic order. Under the perpendicular magnetic field, the spins will tilt and eventually the order and the insulating behavior will disappear.  At much larger fields the bands will be fully split and at $n=1$ an insulating behavior will reappear.

We now attempt a more detailed comparison. In the experimental results, near the boundary of the insulating state a $T$-linear resistivity is reported, with a slope of $\approx 0.06-0.1~\mathrm{k\Omega}$/K (Fig.3 in Ref.~\cite{ghiotto2021quantum}), which corresponds to $5-10\mathrm{k\Omega}/t$ for $t\sim 10 meV \sim 100$K. And the resistivity amplitude is on the order of $\sim4$ k$\Omega$ at 50K. Comparison to our calculated resistivity in Fig.\ref{para_res}(a) indicates that these data correspond to a relatively small $U\sim 4t$. The temperature below which the insulating behavior onsets is $\sim 10$K also consistent with a $U/t \lesssim 4t$. We then fix the interaction at $U=4t$ and plot the phase diagram and the gap size $\Delta$ as a function of chemical potential and $\phi$, as shown in Fig.~\ref{phaseWSe2} (b-d). Experimentally the system is tuned by varying top and bottom gate voltages whose relation to the displacement field and carrier concentration are approximately but not precisely known. Further, for experimental reasons much of the data are obtained by fixing the voltage on the top gate and varying the bottom gate, implying that the carrier concentration and displacement field are simultaneously varied. Our phase diagram (Fig. ~\ref{phaseWSe2}(b)) is roughly consistent with the phase diagram shown in Fig.~1(f) in Ref.~\cite{Wang:2020us} and Fig.~1 (c) of Ref.~\cite{ghiotto2021quantum}, while our Fig. ~\ref{phaseWSe2}(c,d) are consistent with Fig.~1 (d,e) of Ref.~\cite{ghiotto2021quantum}. The maximum reported value of the gap is $3$meV, roughly consistent with out results and the temperature dependence of the resistivity in our Fig. ~\ref{phaseWSe2}(e,f) is similarly  qualitatively consistent with Fig.~3 (c) of Ref.~\cite{ghiotto2021quantum}. 
 
Another important finding in our study is the onset of the strange metal behavior as the system approaches the high order van hove singularity, where the density has a third-power law divergence $\rho(\epsilon) \sim|\epsilon|^{-1 / 3}$. In Fig.~\ref{phaseWSe2} (e-f), we plot the resistivity for different $\phi$ at half filling. As $\phi$ approaches $\pi/6$, the resistivity not only shows the insulating behavoir, but also goes from Fermi liquid to non/marginal Fermi liquid, consistent with Fig.~5 (a) in Ref.~\cite{ghiotto2021quantum}. 

Finally, we summarize different regions in the resistivity, shown in  Fig.~\ref{phaseWSe2} (f): For $T\lesssim 50$ K ($t$ is chosen as 100K), the system goes in to the insulating state, where $\rho$ decreases as $T$ increases; For $50 \lesssim T \lesssim 100$ K, the system enters the bad metal region. As $T$ further increases to around $100 \lesssim T \lesssim 150$ K, the system experiences a crossover, from a low $T$ regime where the resistivity rapidly increases with $T$ to a high $T$ regime where the resistivity is only weakly dependent on $T$. Remarkably, experiments also observe a relatively sharp defined crossover at $T\approx 100K$ (Fig.~3c in Ref.\cite{ghiotto2021quantum}) A noticeable difference is that the experimentally measured resistivity has less $T$ dependence above the high-T crossover  than is theoretically found. It is possible that this difference arises from extra bands in the real system, which begin to contribute to the conductance at higher energy scales.  

\subsection{MoTe\texorpdfstring{$_2$}{2}/WSe\texorpdfstring{$_2$}{2}}

\begin{figure}[htbp]
	\centering
	\includegraphics[width=1.0\linewidth]{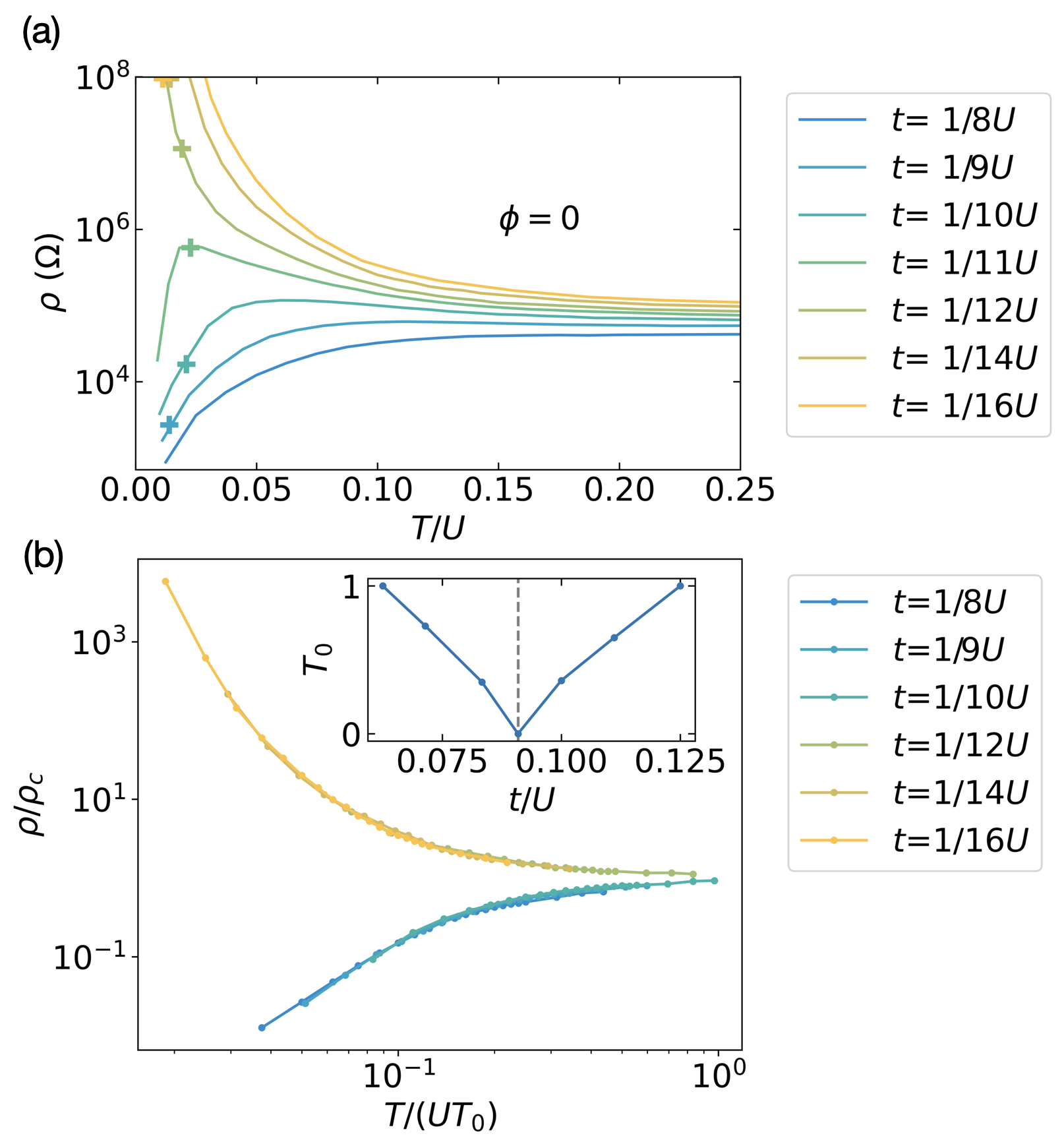}
	\caption{(a): Resistivity near the paramagnetic Mott-insulator transition for $\phi=0$ plotted in log scale with temperature $T$ and hopping $t$ in unit of interaction. The ``plus'' symbol marks the magnetic transition temperature. (b): Resistivity curves collapse onto two branches. The resistivity are scaled by the resistivity at $U=11t$. The temperatures are scaled by a $t$ dependent $T_0$ to collapse the data. The inset shows the value of $T_0$.}
	\label{scale}
\end{figure}
In the moir\'e heterobilayer material MoTe$_2$/WSe$_2$ studied by Li et al \cite{Cornell}, an out-of-plane electrical field $E$ continuously increases the bandwidth while making little change to the interaction strength, inducing a bandwidth driven Mott-insulator transition at half filling consistent with the calculations reported in Fig.~\ref{res0}. The experimentally applied electrical field $E$ is associated with the theoretically defined displacement field $D$ in the corresponding DFT calculation. However, a precise correspondence between the experimentally applied field and the bandwidth is not known, for reasons including uncertainty in the relation between the applied field and the potential drop from one layer to the next, and theoretical uncertainties arising from proximity to a band mixing point at small field and to electric field induced interband transitions at large field \cite{Cornell}.

To make a direct comparison with experiments, we take advantage of the idea that the interaction strength is set by the size of the moir\'e unit cell and is not dependent on the electric field to re-plot our calculated resistivity curves with temperature $T$ and hopping $t$ in units of the interaction $U$, as shown in Fig.~\ref{scale} (a). Our calculation remarkably agrees with the experimental data in Fig.~2(a) of Ref.~\cite{Cornell}. In both the experimental and theoretical data sets the resistivity at the metal-insulator crossover has a peak at a definite temperature. In the experiments the lowest peak is at $T\approx 5$K. We associate this with the $U=11t$ theoretical trace, and then from the peak energy we estimate that the interaction in this system is around 250 K $\approx$ 21.5 meV, with bandwidth$\approx 9t$ changing from 10meV to 25 meV over the relevant range of electric fields ($U/16<t<U/8$). 

We then follow Ref.~\cite{Cornell} and rescale the resistivity data by the ``critical resisitivity'' $\rho_c(T)$ and rescale the temperature by a $t$ dependent parameter $T_0$. As shown in Fig.~\ref{scale} (b), the resistivity curves collapse onto ``insulating'' and ``metallic'' branches. We use resistivity at $t=1/11U$ as $\rho_c$, while $t=1/12U$ would give a similar result. In the scaling process, we use a similar procedure as experiments: we first set $T_0=1$ for the curve furthest from the critical point, and then we determine the value of $T_0$ for the next normalized resistivity curves to best collapse it onto the first curve. Compared with the experimental data (Fig.~2 and 3 of Ref.~\cite{Cornell}), both the collapse and the $t$ dependence of $T_0$ has a very similar behavior. Such scaling is called a ``hidden quantum criticality'' by Dobrosavljevi\'c et al  \cite{Dobrosav_cri,Dobrosav_cri2}. We suggest that the scaling may be understood as follows: single-site DMFT has a $T=0$ metal-insulator transition with a complicated structure involving both the appearance of a mid-gap quasiparticle peak and a gap opening (the two phenomena happen at close but different values of $U/t$). However at high temperatures these two effects are not distinguishable, and one critical behavior associated with a combination of both effects is observed. Consistent with this view is the observation that the scaling behavior is not extremely sensitive to the value of $U_c$ chosen. 

It should be noted that within the DMFT framework, a magnetic transition happens for $U/t\gtrsim 9t$ \cite{wietek2021mott} at low enough temperature, with an estimated $T_c$ marked by the ``plus'' sign in Fig.~\ref{scale} (a). The onset of magnetism is visible within the calculated temperature range, but is not detected in the experiments. It is possible that spatial fluctuation in the moir\'e system suppress the magnetic transition to a lower temperature.

\section{Conclusions }\label{sec:summary}

In this study, we present a comprehensive dynamical mean field study of the moir\'e Hubbard model. We calculate the resistivity over wide temperature and interaction ranges, which provides a direct comparison to the experiments. The new feature in the Hubbard model is the phase $\phi$, which favors the magnetic phase. At $\phi=\pi/6$, the high order van Hove singularity combined with nesting induces magnetic order. Interestingly, this specific type of van Hove with order three ($\rho(\epsilon) \sim|\epsilon|^{-1 / 3}$) also gives a strange metal behavior with a $T$-linear scattering rate and $\omega/T$ scaling. The linear resistivity could survive with non-zero doping. It should be noted that the van Hove singularity of degree three can be realized not just in tWSe$_2$ but also in general moir\'e/multi-layers system \cite{Liang_vhs1,AndreaTrilayer,guerci2021higherorder,shtyk2017electrons}, and may induce interesting behavior. 

Our results elucidate the mechanisms underlying the experimental findings in twisted homobilayer WSe$_2$ and heterobilayer MoTe$_2$/WSe$_2$ \cite{ghiotto2021quantum,Wang:2020us,Cornell}. Both systems exhibit continuous Mott transitions and quantum criticality. We find that the behavior of both systems is well captured by the moir\'e Hubbard model, but in different ways. In tWSe$_2$, the continuous metal-insulator transition is driven by a magnetic transition associated with a change of the hopping phase $\phi$ that brings the high order van Hove point of degree three to the Fermi level. The proximity to this van Hove point also induces a linear resistivity. In MoTe$_2$/WSe$_2$, one has a paramagnetic metal to paramagnetic Mott insulator transition driven by variation of the bandwidth, with the displacement field ($\phi$) effects being 
unimportant. The close correspondence between the data and the calculations reported here establishes the moir\'e Hubbard model as a good theoretical description for wide ranges of temperature and carrier concentrations not too far from the half filled point and motivates further examination of the properties of the model, using methods that go beyond the single site dynamical mean field techniques used here, as well as further experiments to search, among other things, for signatures of the magnetic order found theoretically. 

\begin{acknowledgments}
We thank Augusto Ghiotto, Larry Song, Abhay Pasupathy, and Cory Dean for discussions on experimental results and acknowledge Alexander Wietek and Daniele Guerci for fruitful discussions on the theory. J.Z. thanks Alexander Hampel, Sophie Beck, and Nils Wentzell for valuable discussions on TRIQS software. J.C., J.Z. and A.J.M acknowledge support from the NSF MRSEC program through the Center for Precision-Assembled Quantum Materials (PAQM) - DMR-2011738. The Flatiron Institute is a division of the Simons Foundation.
\end{acknowledgments}

\appendix
\section{band structure of tWSe\texorpdfstring{$_2$}{2}}
\label{appendixA}
The band structure of tWSe$_2$ has been calculated using DFT and continuum theory extensively in previous studies \cite{Wang:2020us,wu2019topological,Sarma,ZangHF}. In this section, we briefly review the band structure and explain how the moir\'e Hubbard model in Eq.~\ref{eq:H} captures the most important features relevant to the physics discussed in this paper.
\begin{figure}[htbp]
	\centering
	\includegraphics[width=1.0\linewidth]{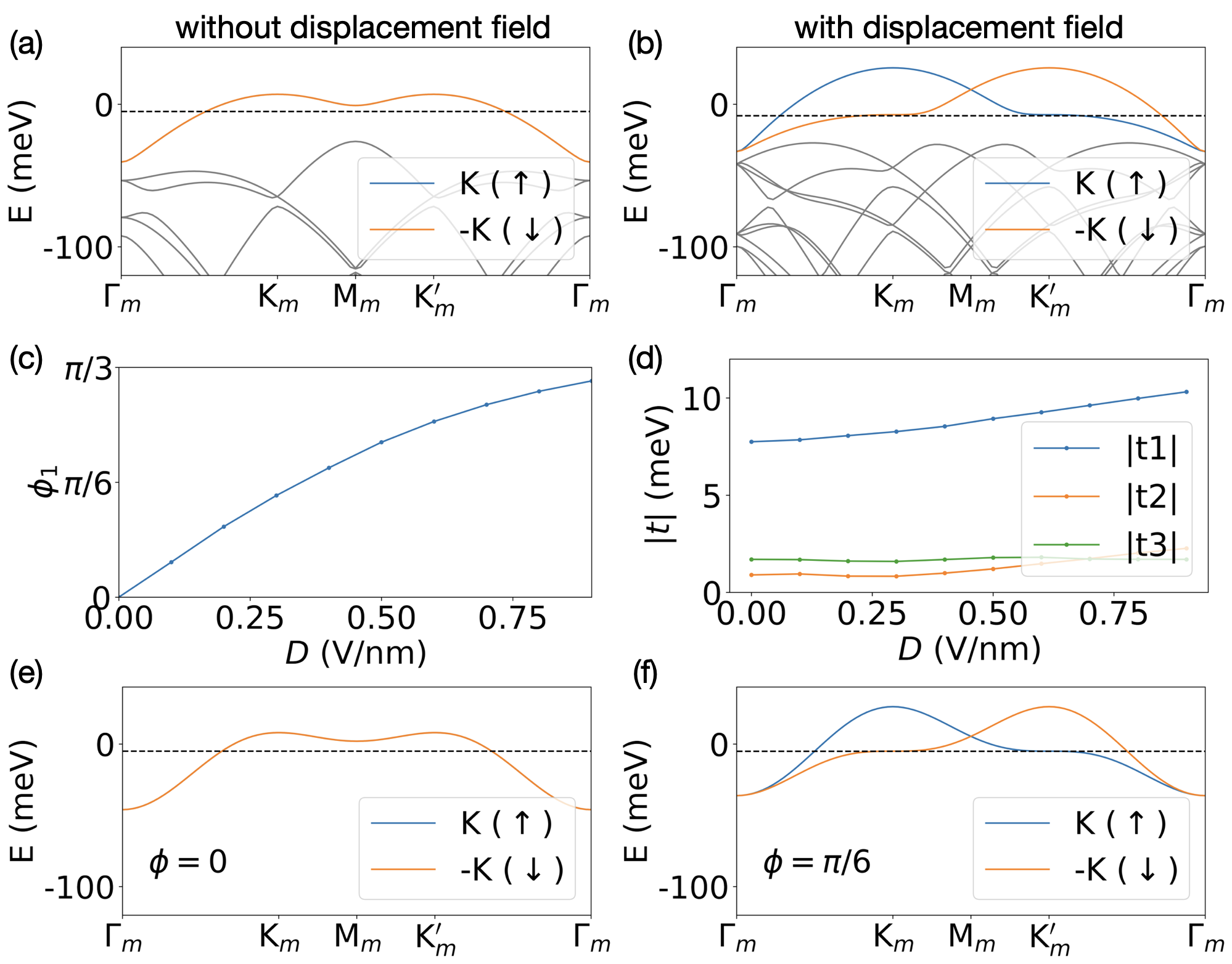}
	\caption{(a,b) Sketch of the band structure of tWSe$_2$ with and without displacement field using continuum model (Refs.[8–10]). The blue (orange) line highlights the topmost valence band from $\vec{K}$ (-$\vec{K}$) valley with spin up (down) dominant. The dotted line indicates the Fermi level at half filling. (c-d) Tight binding parameters as a function of displacement field $D$. The phase of nearest neighbor hopping $\phi_1$ and the hopping amplitude of nearest, second and third neighbors $|t_1|,~|t_2|,~|t_3|$ are extracted from previous DFT and tight binding calculation at $\theta=5.09^\circ$ \cite{Wang:2020us}. (e-f) Band structure of moir\'e Hubbard model (Eq.~\ref{eq:H}) at $\phi=0$ and $\pi/6$, where only the nearest neighbor hopping $t_1$ is considered.}
	\label{band}
\end{figure}

In the monolayer WSe$_2$, the top valence bands at two valleys $\vec{K}$ and $-\vec{K}$ are dominated by opposite spins due to spin oribtal coupling and are related by time reversal symmetry \cite{monolayer}. Stacking two layers of WSe$_2$ with a relative twist angle between them creates an enlarged moir\'e lattice, and a ``flat band'' is formed due to the tunnelling between two layers \cite{Wang:2020us,ghiotto2021quantum,wu2019topological,wu2018hubbard,Sarma}. Owing to the strong spin-orbital coupling in the monolyer, the flatband around valley $\vec{K}$ and $-\vec{K}$ are also dominated by opposite spins. In Fig.~\ref{band}(a,b), we show the band structure calculated from the continuum model \cite{wu2019topological,wu2018hubbard,Sarma} at twist angle $\theta=4^\circ$. For twist angle $4\sim 5 ^\circ$ reported in experiments \cite{Wang:2020us,ghiotto2021quantum}, the top most bands slightly overlap with lower bands at the very lowest band energies according to DFT calculation \cite{Wang:2020us} (also illustrated in Fig.~\ref{band} (a,b)). For density not far below the half filling, the single band description is valid. In our study we consider density around half filling. However, the effect of the hybridization of lower bands should be considered when the density is small. 

Two bands from two valleys are related by time reversal symmetry. For each valley, the top most band can be viewed approximately as a result of hybridization of two parabolic bands $-k^2/2m_{WSe2}$ centered at top-layer $\vec{K}_m$ point and bottom-layer $\vec{K}\prime_m$ point in the moir\'e Brillouin zone, as shown in Fig.~\ref{band}(a). At the high symmetry $\vec{K}_m$ and $\vec{K}\prime_m$ point each spin state comes only from one layer but at a general $\vec{k}$ point each spin state is a combination of the top and bottom layer states. The displacement field changes the relative potential of the top and bottom layers and change the shape of the band structure, as shown in Fig.~\ref{band}(b). 

Previous studies have shown that the topmost energy band can be well described by a tight-binding model on the moir\'e triangular lattice with the phase $\phi$ tuned by the displacement field \cite{Wang:2020us,Sarma,ZangHF}. In Fig.~\ref{band} (c,d), we show the reported tight binding parameters fitted from DFT calculations at twist angle $\theta=5.09^\circ$ \cite{Wang:2020us}. As the displacement field increases, the amplitude of the first nearest neighbor hopping $t_1$ is on the order of $10$ meV, with second and third neighbor hopping smaller than $\sim 20\%$ of $t_1$. The phase $\phi$ of $t_1$ increases from 0 to $\sim\pi/3$ as displacement field increases, as shown in Fig.~\ref{band} (c). The model in the main text (Eq.~\ref{eq:H}) is a simplification of the original tight binding model, where only first neighbor hopping and its phase $\phi$ are retained. In Fig.~\ref{band} (e,f), we plot the band structure of Eq.~\ref{eq:H} at $\phi=0$ and $\pi/6$. Compared to Fig.~\ref{band} (a,b), we can see this model indeed captures the most important features of the effect of the displacement field: the phase $\phi$ generates an inequivalence at moir\'e $\vec{K}_m$ and $\vec{K}_m^\prime$ point, which captures the change of the relative potential of two layers induced by the displacement field. Adding further neighbor hopping will slightly modify the band structure and changes the transition temperature of the magnetic insulator, while it doesn't change the essential physics. 

For this model we consider the onsite interaction only. The nonlocal interaction plays an important role at fractional fillings for heterobilayers, where a Wigner crystal can be realized \cite{CornellWigner,HarvardWigner}. For homobilayer tWSe$_2$ with twisted angle $4\sim 5 ^\circ$, the onsite interaction is not big ($U\sim 4t$), as discussed in the main text. The nearest neighbor interaction of moir\'e system is generally smaller than $20\%$ of onsite interaction from previous estimation \cite{moralesduran2021nonlocal}. Therefore we expect that the nonlocal interaction will not have a strong effect on tWSe$_2$ for the relevant twist angle. Currently no Wigner crystal state has been reported in twisted homobilayer system, consistent with the estimation. It is an open question for investigating the effect of the nearest neighbor interaction in homobilayer system with smaller twisted angle.

Previous studies have also shown that the moir\'e band of tWSe$_2$ can go from topologically trivial to nontrivial region by tuning the parameters of the continuum model \cite{wu2018hubbard, Sarma}. Currently no topological features have been found in experiments and the top most band is well fitted by a single orbital from previous tight binding and DFT calculation at twisted angle $4\sim 5 ^\circ$ \cite{Wang:2020us}. Therefore in this study, we focus on the topological trivial region to study the properties of strange metal and quantum criticality. The study of topology \cite{Sarma,Futopological,wu2019topological} in this system is an important question for future research.

\section{Comparison between \emph{MaxEnt} and Pad\'e methods}\label{app:Pade}
\begin{figure}[htbp]
	\centering
	\includegraphics[width=1.0\linewidth]{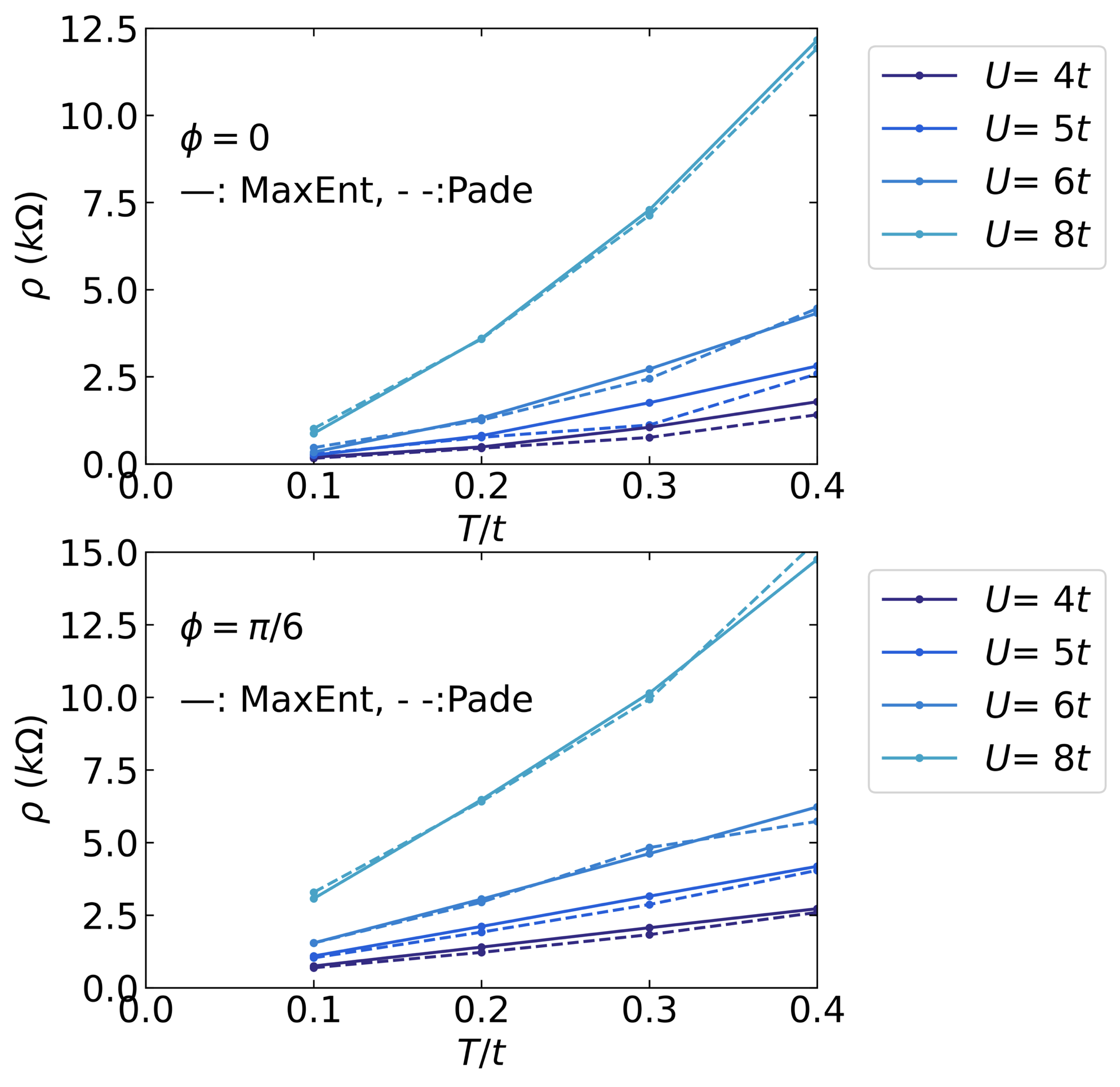}
	\caption{Resistivity of the paramagnetic state for $\phi=0,\pi/6$ at half filling using analytical continuation \emph{MaxEnt} (solid line) and Pad\'e (dashed line) methods.}
	\label{comparison}
\end{figure}
In the main text we mainly use \emph{MaxEnt} continuation method to obtain our resistivity data. Here we perform the continuation via Pad\'e method and compare the resistivity data of the two methods. The continuations for different interactions are performed using the same settings of parameters without fine tuning. As shown in Fig.~\ref{comparison}, results from Pad\'e and \emph{MaxEnt} methods agree well with each other except for $T=0.3t$, where Pad\'e's results deviate a little due to the singularity issues. Generally Pad\'e method is believed to provide a more accurate estimate than \emph{MaxEnt} at low temperature and is less accurate at high temperature. However sometimes Pad\'e has singularity issues. Therefore, in the main text we use \emph{MaxEnt} method to obtain the resistivity data over wide range of temperatures.

\nocite{*}

\bibliography{apssamp}

\end{document}